# Growth methods of c-axis oriented $MgB_2$ thin films by pulsed laser deposition

V. Ferrando [a], S. Amoruso [b], E. Bellingeri [a], R. Bruzzese [c], P. Manfrinetti [d], D. Marrè [a], R. Velotta [c], X. Wang [c], C. Ferdeghini [a]

[a] INFM-LAMIA, Dipartimento di Fisica, Università di Genova, via Dodecaneso 33, 16146 Genova Italy

[b] INFM-Coherentia and Dipartimento di Ingegneria e Fisica dell'Ambiente, Università della Basilicata, C.da Macchia Romana, I-85100 Potenza, Italy

[c] INFM-Coherentia, Dipartimento di Scienze Fisiche, Università di Napoli, Via Cintia, I-80126 Napoli, Italy

[d] INFM, Dipartimento di Chimica e Chimica Industriale, via Dodecaneso 31, 16146 Genova Italy

**Abstract**

High quality $MgB_2$ thin films have been obtained by pulsed laser deposition both on MgO and on $Al_2O_3$ substrates using different methods. In the standard two-step procedure, an amorphous precursor layer is deposited at room temperature starting both from stoichiometric target and from boron target: after this first step, it is annealed in magnesium atmosphere in order to crystallize the superconducting phase. The so obtained films show a strong c-axis orientation, evidenced by XRD analysis, a critical temperature up to 38 K and very high critical fields along the basal planes, up to 22T at 15K. Also an in situ one step technique for the realization of superconducting $MgB_2$ thin films has been developed. In this case, the presence of an argon buffer gas during deposition is crucial and we observe a strong dependence of the quality of the deposited film on the background gas pressure. The influence of the Ar atmosphere has been confirmed by time and space-resolved spectroscopy measurements on the emission spectrum of the plume. The Ar pressure modifies strongly the plasma kinetics by promoting excitation and ionization of the plume species, especially of the most volatile Mg atoms, increasing their internal energy.

**Introduction**

The main problem in $MgB_2$ thin films deposition is the high volatility of Magnesium which rules out the use of high substrate temperatures necessary for film deposition. There are two different approaches to take into account this problem. The first one, also used in bulk preparation, consists in producing a Mg atmosphere by putting some Mg in a sealed container, so that high temperatures can be used. The second one consists in depositing the film at very low temperatures in order to reduce magnesium losses. In the case of thin films preparation, the first approach needs two steps [1]: the deposition of a non superconducting amorphous precursor layer (B or Mg-B) and the heating treatment at high temperature to obtain the superconducting phase. However, for electronic applications that require the integration of different materials as in devices or junctions, the annealing procedures present some drawbacks. Therefore, an in situ deposition of superconducting $MgB_2$ seems necessary. Anyway, as grown synthesis is quite rare in literature and only few groups succeeded in this kind of $MgB_2$ thin films deposition [2-5] using different techniques as MBE, Sputtering and Pulsed Laser Deposition (PLD). In all cases, the quality of the deposited films is lower than that obtained by the two step methods. A complete review of thin films deposition appeared recently in the literature [6].

In this paper we present results in thin films deposition by both methods. In the first section we describe the single step method for as grown superconducting thin film deposition by pulsed laser ablation. Through optical spectroscopy measurements we clarify the crucial role of the argon buffer atmosphere to reach the favourable conditions for the deposition in the plasma plume. In the second section the results obtained with the two step procedure, as well as the superconducting, morphological and structural characteristics of the films are presented.

**1. As grown films**

As grown $MgB_2$ superconducting thin films have been prepared by PLD technique on Sapphire and MgO substrates[2]. The crucial condition to have superconducting films is the use of an Argon buffer gas in the deposition chamber, that changes the dynamics of the species in the plasma plume during its expansion. In fact, only the samples grown in a small range of Ar pressures ($10^{-2}$-$10^{-1}$ mbar) turned out to be superconducting. The first attempts of thin film growth by this route have been performed starting from a Mg enriched target, prepared by mixing and pressing Magnesium and Boron powders in a ratio of 1:1, without any thermal reaction. This was done in order to mitigate Mg losses caused by the high temperature. Samples grown from this kind of target, in a temperature range from 400 to 450°C, showed superconducting properties, even though with a low critical temperature of about 25 K. The Ar pressure was about $10^{-2}$ mbar. Recently, we observed superconductivity with a similar value of $T_c$ starting from stoichiometric $MgB_2$ sintered target. In this case, the Ar pressure was about $10^{-1}$ mbar and the use of a lower substrate temperature (around 300-350 °C) was needed to avoid magnesium evaporation from the films. As explained in what follows, the argon pressure is a critical parameter for this single step procedure, the quality of the sample being strongly dependent on its value. In particular, the Ar pressure strongly influences the colour of the plume emission, which passes from green to blue when the Ar pressure is increased. According to our experience, optimal conditions for film deposition are only achieved when the plume emission is blue.

In order to clarify how the argon pressure influences the plume characteristics and to find the best condition for the film growth, we have carried out an optical emission investigation of the plume produced by excimer laser ablation of the $MgB_2$ expanding in different Ar buffer pressures. The measurements have been performed in conditions similar to the ones used for the films deposition. Time integrated and time- and space-resolved studies of the strongest emission lines of neutral and ionic excited species of the laser ablated plasma over a broad wavelength range (200-800 nm) have been performed. In particular, our analysis has allowed us to relate the change in the plasma emission, above a specific Ar pressure, to the onset of a blast wave in the plume expansion into the buffer gas.

The bright plasma emission was viewed through a side window at right angles to the plume expansion direction. A slice of the plasma was imaged onto the entrance slit of a 0.25 m monochromator equipped with either a 1200 or 100 grooves/mm grating (maximum resolution ≈0.05 nm and ≈ 0.5 nm, respectively). The monochromator exit was coupled to either an intensified charged coupled device (ICCD) camera, with a minimum temporal gate of 5 ns and an overall spectral resolution of about 0.4 nm, for time-integrated measurements, or to a photomultiplier tube to trace the temporal evolution of selected spectral lines of Mg and B excited atoms and ions.

We report in figure 1 emission spectra from the $MgB_2$ expanding plume at different Ar background pressures for a representative distance $d = 5$ mm from the target surface. The identification of Mg and B lines was accomplished by standard data available in the literature [7]. The broad peak in the UV (around 285 nm) in fig. 1 is due to the overlapping of spectrally very close emission lines of excited B and Mg atoms and ions, unresolved by the 100 grooves/mm grating.

The spectra mainly consist of neutral B and Mg lines (B I and Mg I), with some contributions from their excited ions (B II and Mg II), and the plume emission in the visible range can be mainly ascribed to Mg I and Mg II, since the most intense B lines belong to the UV spectral range. A first interesting aspect is that no intense molecular bands are observed in the emission spectrum, indicating a scarce presence of molecular compounds in the plume.

A second prominent feature of the spectra of fig. 1 is that while at low buffer pressure (up to $\approx 10^{-1}$ mbar) the contribution from the Mg I 518 nm green line to the emitted visible light is larger then that of the Mg II blue line at 448 nm, this last line strongly increases as the pressure grows up, becoming almost an order of magnitude larger than the green line at the larger pressures. Thus, the presence of an Ar buffer leads to a variation in the plume colour and shape, as also reported in refs. 2 and 3: in our experimental conditions, a green plume is observed close to the target for an Ar pressure up to $\approx 7 \times 10^{-2}$ mbar, while above this value a long and bright sky blue plasma plume is formed. Then, the difference in the plume emission observed at different Ar background pressures can be totally ascribed to the excitation and ionisation dynamics of the different species contributing to the visible plume. This makes questionable the influence of oxygen impurities, possibly present in the target, or of molecular oxides of the sample atoms on plume emission, as recently suggested in the literature [8].

Time-integrated measurements have also showed that emission intensity of the three representative Mg atom and ion lines of fig. 1 is strongly dependent on the Ar background pressure. The general trend is that of a strong decrease of the photon yield up to an Ar pressure of $\approx 7 \times 10^{-2}$ mbar, followed by a steep increase at higher pressures. This evidences the central role played by the Ar buffer gas in determining the plasma kinetics by promoting excitation and ionisation of the plume species and, in particular, of Mg atoms.

In order to confirm the influence of the buffer gas on plume dynamics, the evolution of selected species has been studied by time-resolved optical spectroscopy. The main outcomes of this analysis are: a) the appearance of a long tail in the optical time-of-flight distributions of all the emission lines, and a strong decrease of atoms velocity for pressures above $10^{-1}$ mbar, indicating the occurrence of a braking effect due to collisions of the expanding plume particles with the buffer gas; b) the enhancement of the plume excitation and ionisation degree at the same buffer gas pressure.

A detailed analysis of both the dependence of the plume expansion velocity on the distance from the target, and of the plume temperature on the buffer gas pressure, through Boltzmann plot method, has lead us to interpret our results in terms of the formation of a blast wave during the plume expansion into the buffer gas [9]. In fact, as already discussed in the literature [10,11], the occurrence of a shock wave causes the redistribution of kinetic and thermal energies between the plume and the ambient gas, leading to a significant transformation of the particles flux energy into plume thermal energy.

Since the excitation state and the kinetic energy of an impinging atom greatly influence the process of thin films formation [12], the mechanism we have observed lends itself as a powerful means for the control of $MgB_2$ film growth. In fact energetic particles cause re-sputtering and largely influence the deposit content especially of the most volatile elements, like Mg in $MgB_2$. Thus, a decrease in the energy of the impinging species could contribute to the increase of the Mg content in the deposit. Moreover, the arrival of a larger extent of excited species on the growing film produces an increase of the energy released in the impact region, thus enhancing surface atomic mobility and favouring film growth at lower substrate temperatures. As a consequence, one can select appropriate buffer gas pressure and target-to-substrate distances to optimise the amount of Mg plume atoms with a higher degree of internal excitation energy and a sufficiently low kinetic energy impinging on the substrate. This deposition method even if still not fully optimised is promising for realizing $MgB_2$ based devices: this study will be greatly helpful to deposit via laser ablation as-grown, superconducting thin films of $MgB_2$.

**2. Two-step synthesis**

Although single step deposition of MgB$_2$ thin films is very important for electronic applications, the quality of the films obtained so far is not satisfactory, and the process is not completely reproducible. On the contrary, the two step procedure, firstly proposed in ref. [1], has led to better results in the deposition of high quality MgB$_2$ thin films.

The first step consists in the PLD deposition of a precursor layer in UHV conditions (10$^{-9}$ mbar) at room temperature. This precursor layer can be obtained both from a stoichiometric MgB$_2$ sintered target and from a pure boron target (prepared by pressing amorphous boron powders). These layers are amorphous and not superconducting. Therefore, an ex-situ annealing process in magnesium atmosphere is needed to crystallise the superconducting phase. For this purpose, the samples are sealed in tantalum crucibles under argon pressure with Mg lumps, closed in an evacuated quartz tube and kept at 850 °C for half an hour with a following rapid quenching to room temperature. Due to the high reaction temperatures involved in this process, we chose MgO and Al$_2$O$_3$ as substrates for their thermal stability during the heat treatment.

In particular, the use of MgO (111) and sapphire c-cut is particularly favourable, due to their hexagonal surface symmetry, which is similar to that of MgB$_2$. This has indeed led to in plane oriented growth of the samples. Thin films have been characterized both structurally, by standard x-rays diffraction and synchrotron radiation (ID32 Grenoble) [13], and morphologically, by AFM. Resistivity down to 4.2 K has been measured by a standard four probe technique.

Despite films grown from Boron precursors show high critical temperatures (close to the bulk values), differently from the results presented in ref[1], their structural characteristics are poor. In the θ-2θ X-rays pattern in fact, only very low intensity (00l) reflections have been detected. On the contrary, films grown from MgB$_2$ precursors showed good structural properties even though with low T$_C$ values. In both cases the presence of (00l) reflections of MgB$_2$ is evident and the (101) peak, the most intense in powders, has lower intensity or is totally absent, indicating a strong c-axis orientation. An example of XRD spectrum of a film deposited on Al$_2$O$_3$ c-cut single crystal substrate starting from stoichiometric target is reported in fig.2. It is important to underline that in this case the (101) peak is absent while spurious reflections, probably arising from air contamination, are visible. The strong c-axis orientation is confirmed also by the Rocking Curve measurement around the (002) reflection, which is reported in Fig 3a. The curve shows a good gaussian shape with a FWHM of only 0.63° indicating a very good alignment of the c-oriented grains. However, it must be noted that the curve intensity background indicates that a randomly oriented part of the MgB$_2$ phase is still present.

The epitaxiality of MgB$_2$ thin films is still a quite difficult task, and only few groups succeeded in producing in plane oriented samples [14,15]. From φ-scan measurements on a four circle diffractometer, we obtained indication of in plane orientation in the case of the film of fig2. An example of this measurement performed on the (101) MgB$_2$ reflection (2θ=42.6) is reported in fig 3b for φ in the range 0–180°. Two families of peaks can be observed: two high and narrow peaks spaced by 120° belonging to the (113) reflection of the Al$_2$O$_3$ substrate (3-fold symmetry) and three lower, larger peaks, spaced by 60°, which are the (101) of MgB$_2$ (six fold symmetry axes). From the analysis of these measurements, we can conclude that the MgB$_2$ is strongly c-axis oriented and that these c axis oriented grains grow epitaxially on the substrate with the a axis rotated of 30° with respect to the a axis of the sapphire. This rotation of 30° with respect to the substrate lattice can be explained by the lower lattice mismatch between MgB$_2$ and substrate along this direction (12% instead of 36% for the *a* to *a* alignment).

From an electrical point of view, the films prepared from stoichiometric target show critical temperature slightly lower than the bulk value, up to 35 K, and also lower residual resistivity ratios (up to 1.6) than those prepared from boron precursors.

Electrical resistance was measured as a function of temperature in applied magnetic field up to 9T: the magnetic field was applied parallel and perpendicular to the film surface and the current was always perpendicular to the magnetic field. From these measurements we estimated H$_{c2}$ for each temperature as the point of the transition in which the resistance is the 90% of the normal state value. In figure 4 critical fields in the two orientations (parallel and perpendicular to the basal plane) for four different films are presented. The films are ordered from left to right increasing critical temperature and RRR. Some structural and electrical characteristics of the four samples are summarized in Tab. 1.

For the film with the lowest $T_C$ and RRR values, a linear $H_{c2}$ versus temperature dependence is evident, while for the other three samples an upward curvature near $T_C$ (that has been related to the clean limit) appears, becoming more and more evident increasing $T_C$ and RRR.

One of the topics still not clarified in $MgB_2$ phenomenology is the anisotropy value. In fact, significant differences in γ values are reported in literature, depending on the kind of the sample and on the measuring method. The anisotropy factor is usually defined as the ratio between the upper critical fields parallel and perpendicular to the basal planes. Due to the multiband nature of superconductivity in this compound, the relation between γ and the effective masses ratio is not simple and the anisotropic Ginzburg-Landau model cannot be trivially applied. The γ values reported in literature for single crystals (ranging from 2.6 up to 6 )[16-19] are, in general, higher than those measured on c-oriented thin films, and present a stronger temperature dependence. In the case of thin films, where the effect of disorder is important, γ shows lower values and ranges between 1.2 and 2.4 [13, 20-23].

From the $H_{c2}$ values at lowest temperatures reported in fig.4, it is possible to calculate the anisotropy factor γ that resulted to be 3.0 for film 1, 1.95 for film 2, 1.25 for film 3 and 1.15 for film 4. These values are considerably lower than those of single crystal, and this can be ascribed to incomplete film orientation. With the purpose to roughly compare the different texturing of our films, we can define a texturing coefficient $t = \left[\dfrac{I^{film}(002)}{I^{film}(101)}\right]\left[\dfrac{I^{powder}(101)}{I^{powder}(002)}\right]$ where $I^{film}(002)$ and $I^{film}(101)$ are the measured intensities of the (002) and (101) reflection and $I^{powder}(002)$ and $I^{powder}(101)$ are the tabulated intensities of randomly oriented powders. In table I, the texturing coefficients for the films are reported. We must consider that films 3 and 4 resulted to be more oriented than film 2, while they present lower γ. Therefore the difference in γ among the four films cannot be only ascribed to the different grade of texturing. In any case, these values lie in the literature range despite the different substrates used. We believe that, when $T_C$ increases, also the RRR factor increases, indicating a passage between dirty and clean limit conditions. While in the dirty limit case we observe high anisotropy values, nearly temperature independent, in the clean limit condition a lower (at least near $T_C$), temperature dependent anisotropy is obtained. The boron precursor method more easily produces samples with higher $T_C$ and RRR values and, therefore, clean limit conditions [21].

In general thin films show also higher $H_{c2}$ values with respect to single crystals [24]. This is evident in fig. 5, where the critical fields in both directions are reported for film1. It must be noted that in this case the standard BCS formula, which estimates $H_{c2}(0)$ from the slope of the curve near $T_C$, fails. In fact, in the case of the magnetic field parallel to the film surface, the BCS extrapolation value of 22 Tesla at 0 K is already reached at 13K. The linear temperature dependence is present even at the lowest temperatures we measured.

## 3. Conclusions

In this paper we have described two different procedures to prepare $MgB_2$ thin films by PLD. The first one is an in situ method based on the use of Ar atmosphere in the deposition chamber, that allows to obtain as grown superconducting thin films, even though of poor electrical and structural properties. Emission optical spectroscopy studies of the plume have allowed us to clarify the crucial role of the argon buffer gas pressure in determining the correct conditions for PLD. In the second procedure, which is a two step method, the growth of high quality thin films is possible. The obtained samples showed superconducting properties (critical temperatures, critical fields, RRR and anisotropies) different from bulk single crystals. In this kind of MgB$_2$ thin films, epitaxial growth was obtained even though associate to low critical temperature, using sapphire c-cut as a substrate and stochiometric MgB$_2$ as a precursor layer.

**Acknowledgments**

This work is partially supported by the Istituto Nazionale per la Fisica della Materia through the PRA UMBRA. We acknowledge also the support of the European Community trough "Access to Research Infrastructure action of the Improving Human Potential Programme ".

**Figure captions**

Fig1. Optical emission spectra of the $MgB_2$ plume at different Ar buffer pressures, for d = 5 mm. The most intense lines are labelled in the figure.

Fig.2 XRD ($\theta$-$2\theta$) spectrum for $MgB_2$ thin films prepared from stoichiometric target on $Al_2O_3$ c-cut substrate. Despite some spurious peaks, only the (00l) reflections are evident for the $MgB_2$ phase.

Fig3. (a) Rocking curve around the (002) reflection: the FWHM is 0.63°. (b) X-rays $\phi$-scan for the (101) $MgB_2$ reflection; the peaks are spaced by 60°. The $Al_2O_3$ substrate peaks are 30° shifted with respect to the $MgB_2$ ones and spaced by 120°.

Fig.4 $H_{c2}$ versus temperature for four different films (film1,2,3,4 of table I) for magnetic fields parallel (full symbols) and perpendicular (open symbols) to the film surface.

Fig.5 $H_{c2}$ and $H_{irr}$ versus temperature for film1.

| Sample | Film1 | Film2 | Film3 | Film4 |
|---|---|---|---|---|
| Preparation | From MgB$_2$ | From MgB$_2$ | From MgB$_2$ | From Boron |
| Substrate | Al$_2$O$_3$ c-cut | MgO(111) | Al$_2$O$_3$ r-cut | MgO(111) |
| T$_c$ [K] | 29.6 | 31.4 | 36.2 | 37.5 |
| ΔT$_c$ [K] | 2.3 | 1.1 | ~6 | 0.6 |
| RRR | 1.1 | 1.2 | 1.5 | 2.4 |
| Texturing | Not defined, no (101) refl. | 1.6 | 22.6 | 5.6 |
| FWHM (002), [deg] | 0.6 | ~8 | 2 | 3 |
| γ | 3.0 | 1.95 | 1.25 | 1.15 |

Table I. Characteristics of the four films of fig.4.

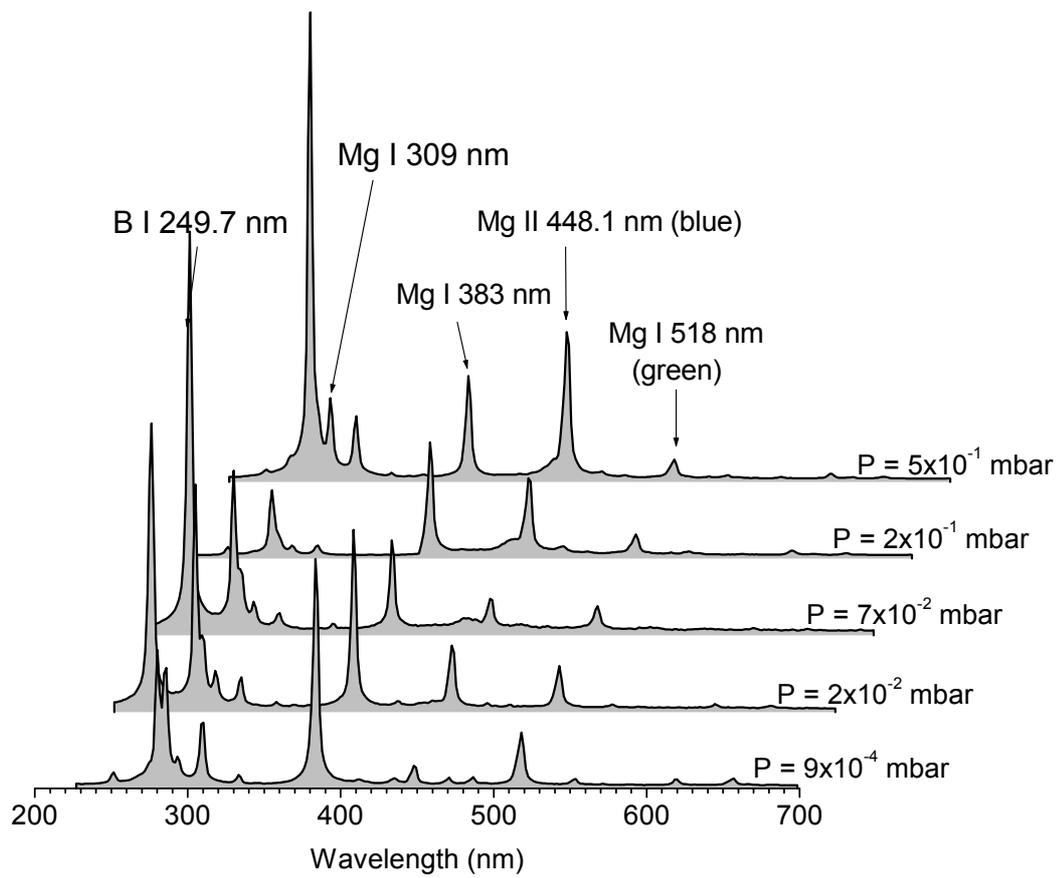

Figure 1

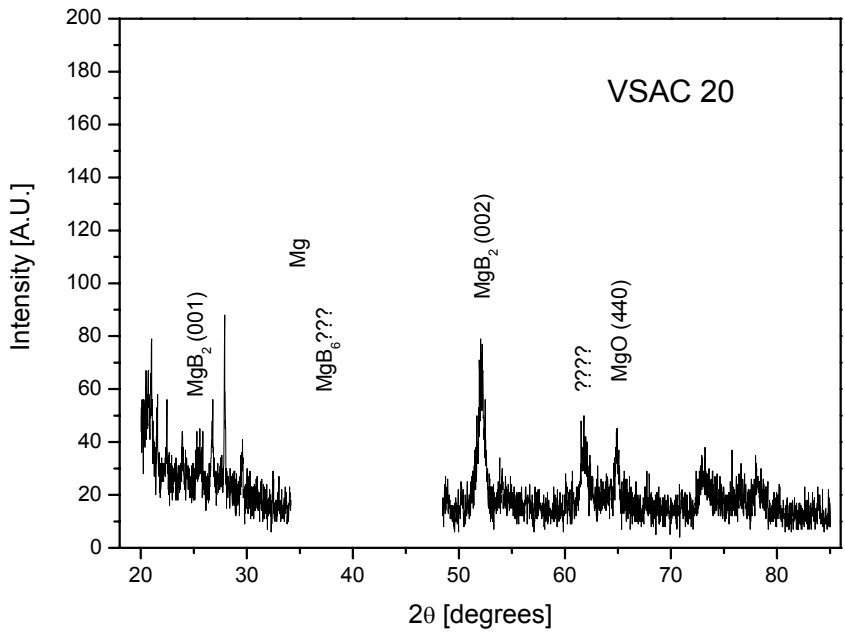

figure 2

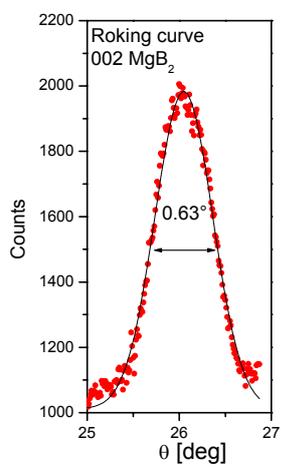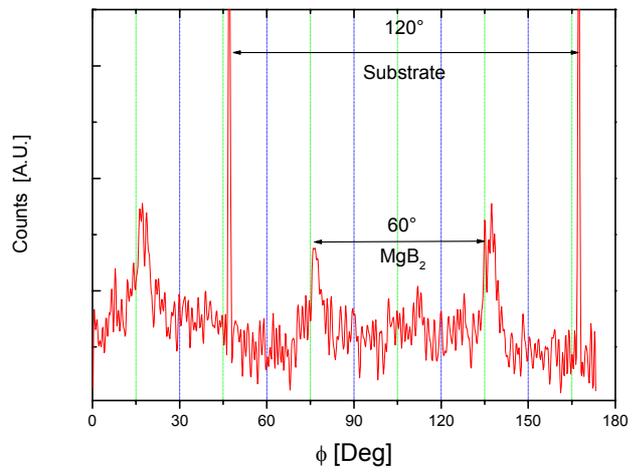

figure 3

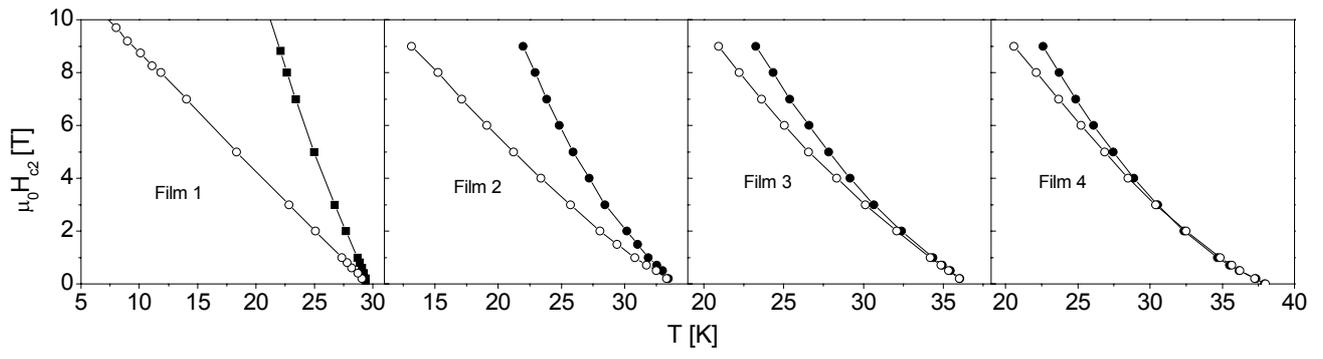

Figure 4

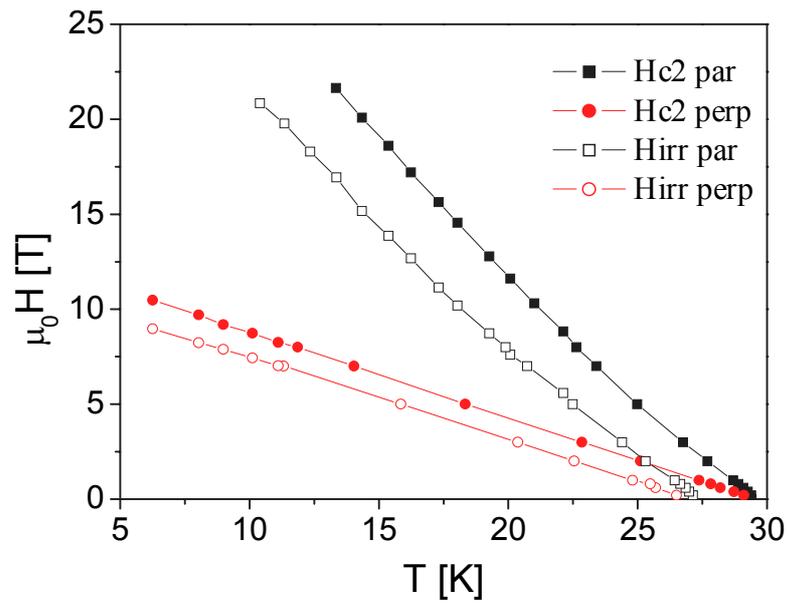

figure 5